\documentstyle[aps]{revtex}
\tighten

\begin{document}


\preprint{YITP-99-75, CITA-99-45, gr-qc/9912xxx}

\title{Excitation of Kaluza-Klein gravitational mode}
\author{Kunihito Uzawa}
\address{
Graduate School of Science and Technology, Chiba University\\
Chiba 263-8522, Japan}
\author{Yoshiyuki Morisawa}
\address{
Yukawa Institute for Theoretical Physics, Kyoto University\\
Kyoto 606-8502, Japan}
\author{Shinji Mukohyama}
\address{
Department of Physics and Astronomy, University of Victoria\\
Victoria BC, Canada V8W 3P6\\
Canadian Institute for Theoretical Astrophysics, 
University of Toronto\\
Toronto ON, Canada M5S 3H8\\
}
\date{\today}

\maketitle


\begin{abstract} 

We investigate excitation of Kaluza-Klein modes due to the parametric 
resonance caused by oscillation of radius of compactification. We
consider a gravitational perturbation around a $D$-dimensional
spacetime, which we compactify on a ($D-4$)-sphere to obtain a
$4$-dimensional theory. The perturbation includes the so-called
Kaluza-Klein modes, which are massive in $4$-dimension, as well as
zero modes, which is massless in $4$-dimension. These modes appear as
scalar, vector and second-rank symmetric tensor fields in the
$4$-dimensional theory. Since Kaluza-Klein modes are troublesome in
cosmology, quanta of these Kaluza-Klein modes should not be excited
abundantly. However, if radius of compactification oscillates, then
masses of Kaluza-Klein modes also oscillate and, thus, parametric
resonance of Kaluza-Klein modes may occur to excite their quanta. In
this paper we consider part of Kaluza-Klein modes which correspond to
massive scalar fields in $4$-dimension and investigate whether quanta
of these modes are excited or not in the so called narrow resonance
regime of the parametric resonance. We conclude that at least in the
narrow resonance regime quanta of these modes are not excited so 
catastrophically. 

\end{abstract}



\section{Introduction}
	\label{sec:introduction}

When we consider unified theories, spacetime may have more than four
dimension. For example, the critical dimension of the superstring
theories~\cite{Superstring} is equal to ten, and the low-energy
effective theory of M-theory~\cite{M-theory} is the eleven-dimensional 
supergravity. On the other hand, we recognize only four dimensional
spacetime. Therefore, in order to describe our universe we have to
adopt some mechanism of dimensional reduction. Compactification of
extra dimensions on some compact manifold (the Kaluza-Klein 
prescription~\cite{KK}) has been conventionally adopted, while an
alternative prescription was recently proposed by Randall and
Sundrum~\cite{RS}. Since the Randall-Sundrum prescription cannot be
applied directly to more than $5$-dimension, in this paper we consider
the Kaluza-Klein prescription.

However, there are too many ways of compactification to specify a
$4$-dimensional theory from the higher dimensional unified
theories. As for the superstring theory, although it is usually
compactified on a Calabi-Yau manifold, there are so many Calabi-Yau
manifolds.

Fortunately, in some cases it may be possible to judge which way of
compactification is acceptable since different compactifications may
give different $4$-dimensional theories and, thus, different
cosmologies. In particular, properties, eg. mass spectrum, of 
the so-called Kaluza-Klein modes reflect the manifold on which
spacetime is compactified. Thus, it may be effective to investigate
effects of Kaluza-Klein modes on the $4$-dimensional
cosmology.

In this respect, Kolb \& Slansky~\cite{Kolb&Slansky} considered the
effects of Kaluza-Klein modes on our $4$-dimensional universe. For
simplicity, they considered a scalar field in $5$-dimensional
spacetime of the form $M^4\times S^1$, where $M^4$ is a
$4$-dimensional FRW universe and $S^1$ is a circle. They investigated
Kaluza-Klein modes of the scalar field and showed that the
Kaluza-Klein modes are dangerous in cosmology. To be precise, they
showed that, if energy density of a Kaluza-Klein mode is not
negligible compared with radiation density at an early epoch, then 
the Kaluza-Klein mode eventually dominate the universe. Thus, quanta
of Kaluza-Klein modes should not be excited abundantly. It is expected
that this result holds in quite general situation since their
arguments are based on the momentum conservation along the compact
direction. Thus, {\it quanta of Kaluza-Klein modes of any fields
should not be excited abundantly}.

However, there is a possibility that quanta of Kaluza-Klein modes may
be excited. First, the radius of compactification appears as a scalar
field in $4$-dimension called a radion~\cite{radion}. Hence, if the
radion potential has a local minimum then it can oscillate around the 
local minimum. The oscillation of the radion leads to oscillation of
masses of Kaluza-Klein modes since the masses are proportional to the
inverse of the compactification radius. On the other hand, it is
well-known that oscillation of mass leads to the so-called parametric
resonance phenomenon~\cite{Landau,Reheating}. Hence, it is expected
that, if the radion potential has a local minimum then quanta of
Kaluza-Klein modes may be excited by parametric
resonance~\cite{Mukohyama}.

Therefore, it is possible to obtain some knowledge about a way of 
compactification by investigating whether quanta of Kaluza-Klein modes
are excited by the parametric resonance due to the radion
oscillation. If the quanta are produced abundantly then the way of
compactification should be rejected. In Ref.~\cite{Mukohyama}, the
parametric resonance of Kaluza-Klein modes of a scalar field was
investigated in the regime where the resonance band is narrow (the
narrow resonance regime). It was shown that, for the case of
compactification by a $d$-dimensional sphere, quanta of the
Kaluza-Klein modes are not overproduced.

However, since there should be many other fields (eg. gravitational
fields, anti-symmetric fields, spinor fields, and so on), the
investigation of only a scalar field is not enough: in principle, 
{\it we have to investigate Kaluza-Klein modes of all fields in the
theory}. In this paper, we investigate a gravitational field.

This paper is organized as follows. In section~\ref{sec:model} we
describe our model and derive the action for our system. In
section~\ref{eqn:resonance} we investigate whether catastrophic
creation of quanta of the Kaluza-Klein modes occur or not. 
In section~\ref{sec:summary} we summarize this paper.


\section{Model construction}
	\label{sec:model}

In this section we derive an effective action for Kaluza-Klein
gravitational modes as well as for background fields. In
subsection~\ref{subsec:assumptions} we describe our model and basic
assumptions verbally. In subsection~\ref{subsection:EHaction} we
calculate the tree action of our system from the Einstein-Hilbert
action. In subsection~\ref{subsec:Casimir} we stabilize the radion
potential by the so-called Casimir effect to obtain the effective
action for our system.

\subsection{Basic assumptions}
	\label{subsec:assumptions}

As already explained in Sec.~\ref{sec:introduction}, our purpose in
this paper is to judge whether a particular model of dimensional
reduction is acceptable or not by investigating parametric excitation
of Kaluza-Klein modes. For this purpose, we have to specify higher
dimensional theory in $D$-dimension somewhat. Moreover, in order to
obtain the $4$-dimensional effective theory from the $D$-dimensional
theory, we have to specify a particular model of dimensional
reduction. Namely, we consider the following setting. 
\begin{itemize}
 \item [(a)] Let us consider a $D$-dimensional theory which includes 
 gravity, $N_s$ scalar fields, $N_d$ Dirac fields, and other
 fields. The gravity, scalar fields and spinor fields are assumed to
 be described by the Einstein-Hilbert action with a cosmological
 constant, the Klein-Gordon action and the Dirac action,
 respectively. There may be interaction among these fields.
 \item [(b)] We adopt the conventional Kaluza-Klein description of 
 compactification: we compactify the $D$-dimensional spacetime on a 
 $d$-dimensional compact manifold ($d=D-4$) to obtain the
 $4$-dimensional spacetime. As the compact manifold we take a
 $d$-dimensional sphere, which we shall denote by $S^d$.
 \item [(c)] We consider the so-called Casimir effect to stabilize a
 radion potential, where the radion is a $4$-dimensional scalar field 
 corresponding to compactification radius~\cite{radion}. In order to
 stabilize the radion potential by the Casimir effect, as shown in
 subsection~\ref{subsec:Casimir}, we have to assume that $d\ge 2$. 
 \item [(d)] We also consider perturbations of $D$-dimensional fields 
 around the background given by the Kaluza-Klein prescription. These 
 perturbations are described as $4$-dimensional massive fields called
 Kaluza-Klein modes. 
\end{itemize}
For this situation, in order to make calculations possible, we assume
the following assumptions. (See the beginning of
section~\ref{eqn:resonance} for further assumptions.)
\begin{itemize}
 \item [(i)] The number $N_s$ (or $N_d$) of scalar fields (or Dirac 
 fields) is sufficiently large compared with the number of other
 fields included in the $D$-dimensional theory. 
 \item [(ii)] The cosmological constant induced in $4$-dimension is
 zero. 
 \item [(iii)] Terms in the $4$-dimensional effective action which are
 order of $O(\varphi^3)$ can be neglected, where $\varphi$ denotes
 Kaluza-Klein modes. 
\end{itemize}

The assumption (i) makes it possible to calculate corrections to the
radion potential due to the Casimir effect. In principle, all fields
included in the $D$-dimensional theory should contribute to the
Casimir effect. However, if $N_s$ or $N_d$ is large enough then 
contributions from other fields are small compared with those from the 
scalar fields and the Dirac fields. Hence, in this limit, the
correction to the radion potential is parameterized by only $N_s$ and
$N_d$. In fact, we can show that, after redefining the $D$-dimensional
cosmological constant, the corrected potential is parameterized by
only two constants: the $D$-dimensional cosmological constant and a
particular combination of $N_s$ and $N_d$.

Therefore, the assumption (i) combined with the assumption (ii)
determines the corrected radion potential uniquely up to only one 
constant. Hence, as we shall show in subsection~\ref{subsec:Casimir}
explicitly, the potential is parameterized only by the present value
$b_0$ of the compactification radius. Since $b_0$ can be eliminated
from all equations of Kaluza-Klein modes by redefinition of
variables, all information about the corrected radion potential can be
obtained.

On the other hand, the assumption (iii) makes it possible to
investigate each Kaluza-Klein mode independently, provided that
kinetic terms and mass terms are properly diagonalized. In this paper, 
we consider only a particular class of Kaluza-Klein modes
corresponding to $D$-dimensional gravitational perturbations. (In
Ref.~\cite{Mukohyama} a class of Kaluza-Klein modes corresponding to
a $D$-dimensional scalar field was considered.) It is the assumption
(iii) that makes it possible to consider these modes independently.

\subsection{Perturbed Einstein-Hilbert action}
	\label{subsection:EHaction}

We consider the $D$-dimensional Einstein-Hilbert action with a
cosmological constant: 
%
\begin{equation}
I_{EH}=\frac{1}{2\bar{\kappa}^2} \int d^Dx
\sqrt{-\bar{g}}(\bar{R}-2\bar{\Lambda}),
\label{eqn;1}
\end{equation}
where $ \bar{\kappa} $ is a positive constant, $\bar{R}$ is the
$D$-dimensional Ricci scalar, and $\bar{\Lambda}$ is the cosmological 
constant. We consider a gravitational perturbation $h_{MN}$ around a
background metric $\bar{g}_{MN}^{(0)}$, which we shall specify below. 
%
\begin{equation}
\bar{g}_{MN} =\bar{g}_{MN}^{(0)}+h_{MN}.
\label{eqn;2}
\end{equation}
By substituting Eq.(\ref{eqn;2}) into Eq.(\ref{eqn;1}), we obtain the
perturbed Einstein-Hilbert action as follows. 
%
\begin{eqnarray}
 I_{EH} & =  & \frac{1}{2\bar{\kappa}^2} 
 	\int d^Dx\sqrt{-\bar{g}^{(0)}} 
	\left[ \bar{R}^{(0)} -2\bar{\Lambda}
	-h^{MN}\left(\bar{R}^{(0)}_{MN}
	-\frac{1}{2}\bar{R}^{(0)}\bar{g}^{(0)}_{MN}
	+\bar{\Lambda} \bar{g}^{(0)}_{MN}\right)\right.\nonumber\\
 & & + \frac{1}{8}\left( h^2-2h^{MN}h_{MN}\right)\bar{R}^{(0)} 
	+\frac{1}{2}\left(2h^{MM'}h_{M'}^N-hh^{MN}\right)
	\bar{R}^{(0)}_{MN}  \nonumber  \\ 
 & &  +\frac{1}{4} \left\{ h^{MN}_{\quad;M'}
	\left(2h^{M'}_{M;N}- h_{MN}^{\quad;M'}\right) 
	+ h_{;M} \left(h^{;M}-2h^{MN}_{\quad;N}\right) 
	\right\}\nonumber\\
 & & \left.
	-\bar{\Lambda}\left(\frac{1}{4}h^2-\frac{1}{2}h^{MN}h_{MN} 
	\right)  +O(h^3)\right].
	\label{eqn:perturbedEH}
\end{eqnarray}
where ``$;$'' denotes the covariant derivative compatible with
$g^{(0)}_{MN}$, and $\bar{R}^{(0)}_{MN}$ and $\bar{R}^{(0)}$ are 
the Ricci tensor and scalar constructed from
$\bar{g}^{(0)}_{MN}$. (See Appendix~\ref{sec:action} for a detailed
derivation.)

As for the background geometry, we compactify it on a $d$-dimensional
sphere $S^d$ ($d=D-4$): 
%
\begin{equation}
\bar{g}^{(0)}_{MN}dx^Mdx^N=\hat{g}_{\mu\nu}dx^{\mu}dx^{\nu}+ 
b^2\Omega_{ij}^{(d)}dx^idx^j, 
\end{equation}
where $ \hat{g}_{\mu\nu} $ and $b$ are a $4$-dimensional metric and a
$4$-dimensional scalar that depend only on the four-dimensional
coordinate $x^{\mu}$ ($\mu=0, 1, 2, 3$), and $\Omega_{ij}$ is a metric
of the unit $d$-sphere which depends only on the coordinates $x^i$
($i=4,\cdots, D-1$). Hereafter, we denote $D$-dimensional,
$4$-dimensional and $d$-dimensional indexes by capital Latin
($M,N,\cdots$), Greek ($\mu,\nu,\cdots$) and small Latin
($i,j,\cdots$) letters, respectively. Note that the scalar $b$ can be 
interpreted as radius of $S^d$.

Since it is convenient to analyze in the so called Einstein frame, we
perform the following conformal transformation to make the new frame
to be the Einstein frame. 
%
\begin{equation}
\hat{g}_{\mu\nu}=\left(\frac{b}{b_0}\right)^{-d} g_{\mu\nu}.
\end{equation}
Note that, if we take $b_0$ to be the present value of $b$, then the
conformal factor becomes unity after $b$ settles to $b_0$. Hence, the
change of the frame does not affect the final result, provided that
all physical results are interpreted after $b$ settles to $b_0$. On
the other hand, our analysis becomes much easier if we adopt this
conformal transformation. After the conformal transformation, the 
$D$-dimensional line element is written as 
%
\begin{equation}
\bar{g}^{(0)}_{MN}dx^Mdx^N=\left(\frac{b}{b_0}\right)^{-d}
g_{\mu\nu}dx^{\mu}dx^{\nu}+ b^2\Omega_{ij}^{(d)}dx^idx^j .
\end{equation}

For this particular background, the gravitational perturbation
$h_{MN}$ can be expanded by harmonics on $S^d$ as follows 
%
\begin{eqnarray}
 h_{MN}dx^Mdx^N &=& \sum_{lm}
 	\left[ h_{\mu\nu}^{lm}
	Y_{lm}dx^{\mu}dx^{\nu}  
 	+2 \{h_{(T)\mu}^{lm}(V_{(T)lm})_i
	+ h_{(L)\mu}^{lm}(V_{(L)lm})_i \}
	dx^{\mu}dx^i \right.\nonumber \\
 & & + \left.
	\{ h_{(T)}^{lm}(T_{(T)lm})_{ij}
	+ h_{(LT)}^{lm}(T_{(LT)lm})_{ij}
	+ h_{(LL)}^{lm}(T_{(LL)lm})_{ij}
	+ h_{(Y)}^{lm}(T_{(Y)lm})_{ij}\}
	dx^idx^j \right],\nonumber\\
\end{eqnarray}
where $Y_{lm}$ is the scalar harmonics function; $V_{(T)lm}$ and
$V_{(L)lm}$ are the vector harmonics; $T_{(T)lm}$, $T_{(LT)lm}$, and 
$T_{(LL)lm}$ are the tensor harmonics. Here, the 
coefficients $h_{\mu\nu}^{lm}$, $h_{(T)\mu}^{lm}$, $h_{(L)\mu}^{lm}$, 
$h_{(T)}^{lm}$, $h_{(LT)}^{lm}$, $h_{(LL)}^{lm}$ and $h_{(Y)}^{lm}$
depend only on the four-dimensional coordinates $x^{\mu}$, while the
harmonics depend only on the coordinates $x^i$ on $S^d$. 
(See Appendix~\ref{sec:harmonics} for definitions and properties of
these harmonics.)

Although this expression of $h_{MN}$ includes many terms, some of them 
represent degrees of freedom of coordinate transformations. In fact,
it is shown in Appendix~\ref{sec:fix} that, after gauge-fixing and
redefining $g_{\mu\nu}$ and $b$, the perturbation $h_{MN}$ can be
expressed as follows.  
%
\begin{equation}
 h_{MN}dx^Mdx^N = \sum_{lm}
	\left[ h_{\mu\nu}^{lm}Y_{lm}dx^{\mu}dx^{\nu}
	+ 2h_{(T)\mu}^{lm}(V_{(T)lm})_idx^{\mu}dx^i
	+ \{ h_{(T)}^{lm}(T_{(T)lm})_{ij}
	+ h_{(Y)}^{lm}(T_{(Y)lm})_{ij}\}dx^idx^j\right],
	\label{eqn:gaugefixedh}
\end{equation}
where the summations are taken over $l\ge 1$ for the scalar and vector 
harmonics, and over $l\ge 2$ for the tensor harmonics.

Finally, by substituting this expression into
Eq.~(\ref{eqn:perturbedEH}), we obtain the following action. 
%
\begin{equation}
 I_{EH} = I^{(0)}+I^{(1)}+I^{(2)}+O(h^3),
	\label{eqn:perturbedEH2}
\end{equation}
where 
%
\begin{eqnarray}
& &I^{(0)}=\int d^4x\sqrt{-{g}^{ (0)}}\left[\frac{1}{2{\kappa}^2}R^{ (0)}
-\frac{1}{2}g^{\mu\nu}\partial_{\mu}\sigma\partial_{\nu}\sigma-U_0(\sigma)
\right], \\
& &I^{(2)}=\sum_{l, m}\int d^4x\sqrt{-{g}^{ (0) }}
\left({\cal L}^{(T)}_{lm}+ {\cal L}^{(V)}_{lm}
+ {\cal L}^{(Y)}_{lm} \right),
\end{eqnarray}
and $I^{(1)}$ is linear in $h$. 
When a total action of the system is considered (see the setting (a)
in subsection~\ref{subsec:assumptions}), $I^{(1)}$ should be canceled 
by other linear terms in the total action because of the equations of
motion. Hence, there is no need to consider $I^{(1)}$. The constant
$\kappa$, the scalar field $\sigma$, the potential $U_0$, and
Lagrangian densities ${\cal L}^{(T,V,Y)}_{lm}$ are defined as
follows. First, the constant $\kappa$ ($>0$) is defined by 
%
\begin{equation}
{\kappa}^2=\frac{{\bar{\kappa}}^2}{2^d\;b^d\pi}. 
\end{equation} 
Next, $\sigma$ is a scalar field defined by 
%
\begin{eqnarray}
& &\sigma={\sigma}_0\ln\left(\frac{b}{b_0}\right),\\
& &\sigma_0=\sqrt{\frac{d(d+2)}{2{\kappa}^2}},
\end{eqnarray}
and is called a radion field~\cite{radion}. The potential
$U_0(\sigma)$ of $\sigma$ is given by 
%
\begin{equation}
U_0(\sigma)=\frac{\bar\Lambda}{{\kappa}^2}e^{-d\frac{\sigma}{\sigma_0}}
-\frac{d(d-1)}{2{\kappa}^2{b_0}^2}e^{-(d+2)\frac{\sigma}{\sigma_0}}.
\end{equation}
Finally, the Lagrangian densities ${\cal L}^{(T,V,Y)}_{lm}$ are given
by 
%
\begin{eqnarray}
{\cal L}^{(T)}_{lm}&=&-\frac{1}{2}e^{-4\frac{\sigma}{\sigma_0}}
g^{\mu\nu}\partial_{\mu}\chi^{lm}\partial_{\nu}\chi^{lm}
-\frac{1}{2}\Lambda^{(T)}_{lm}\chi^{lm}\chi^{lm}, \\
{\cal L}^{(V)}_{lm}&=&{\cal L}^{(V)}_{lm}[h^{lm}_{(T)\;\mu}],\\
{\cal L}^{(Y)}_{lm}&=&{\cal L}^{(Y)}_{lm}[\;h^{lm}_{\mu\nu},
h^{lm}_{(Y)}\;],
\end{eqnarray}
where
%

\begin{eqnarray}
 \Lambda^{(T)}_{lm} & = &
	e^{-(d+6)\frac{\sigma}{\sigma_0}}\{l(l+d-1)+(d-1)(d-2)\}
	b_0^{-2}\nonumber\\
	& &+e^{-4\frac{\sigma}{\sigma_0}}\left[(d+4)\nabla^2
	\left(\frac{\sigma}{\sigma_0}\right)
	-\left\{\frac{d(d+2)}{2}-4\right\}\nabla\;
	\left(\frac{\sigma}{\sigma_0}\right)^2 
	+\left\{R^{(0)}-2e^{-d\frac{\sigma}{\sigma_0}}\bar{\Lambda}\right\}
	 \right],\\
 \chi^{lm} & \equiv & \sqrt{\frac{{b_0}^{-4}}{2^{d+2}\pi{\kappa}^2}}
	\;h^{lm}_{(T)} \qquad (l\ge 2). 
\end{eqnarray}
(Since hereafter we analyze $h_{(T)}^{lm}$ (or $\chi^{lm}$) only, we
have not written down explicit form of ${\cal L}^{(V)}_{lm}$ and
${\cal L}^{(Y)}_{lm}$.) Hence, up to the second order in Kaluza-Klein
modes, $h_{(T)}^{lm}$ (or $\chi^{lm}$) is decoupled from all other
Kaluza-Klein modes. (See the assumption (iii) in
subsection~\ref{subsec:assumptions}.)  In this paper, for simplicity,
we investigate the parametric resonance of $\chi^{lm}$ (or
$h_{(T)}^{lm}$) only. It should be stressed again that it is the
assumption (iii) in subsection~\ref{subsec:assumptions} that makes it
possible to consider these modes independently.

Here, note that the constant $b_0$ is arbitrary at the moment and that
we can eliminate it from all equations. However, in the next
subsection, we chose $b_0$ to be the present value of $b$.

\subsection{Stabilization of the radion potential}
	\label{subsec:Casimir}

Here we assume that the Casimir effect~\cite{Candelas&Weinberg}
stabilizes the radion potential. In other words, we assume that
quantum corrections to $U_0$ stabilize compactification. 
Thanks to the assumptions (i) and (ii) in
subsection~\ref{subsec:assumptions}, we can easily calculate the 
stabilized radion potential, which we shall denote by $U_1$.

First, for all Kaluza-Klein modes of $N_s$ scalar fields and $N_d$
Dirac fields, mass in the Einstein frame is proportional to the common
quantity. 
%
\begin{equation}
M^2\propto b_0^{-2}e^{-(d+2)\sigma/\sigma_0}.
\end{equation}
Hence, because of the assumption (i), we can obtain the following
1-loop effective potential. (See, for example, Ref.~\cite{Mukohyama}
for details.) 
%
\begin{equation}
 V_{1\;loop}(\sigma)=Ce^{-2(d+2)\sigma/\sigma_0},
\end{equation}
where $C$ is a constant determined by $N_s$ and $N_d$. There are
corrections to the $D$-dimensional cosmological constant
$\bar{\Lambda}$, too. However, these corrections can be absorbed in
$\bar{\Lambda}$ by redefining $\bar{\Lambda}$. Hence, the corrected
radion potential $U_1$ is given by 
%
\begin{equation}
U_1(\sigma)=U_0(\sigma)+V_{1\;loop}(\sigma). 
\end{equation}

Next, the assumption (ii) requires that the minimum of $U_1$ should be 
zero. Hence, $\bar{\Lambda}$ is determined as follows. 
%
\begin{equation}
\bar{\Lambda}=\frac{d(d-1)(d+2)}{2(d+4)b_0^2},
\end{equation}
where we have specified the constant $b_0$ as 
%
\begin{equation}
 b_0=\sqrt{\frac{d(d-1)}{(d+4)C\kappa^2}}. 
\end{equation}
Thus, we obtain the following radion
potential~\cite{Potential-U1,Mukohyama}.
%
\begin{equation}
 U_1(\sigma) =
	\alpha\left[\frac{2}{d+2}e^{-2(d+2)\sigma/\sigma_0}
	+e^{-d\sigma/\sigma_0}
	-\frac{d+4}{d+2}e^{-(d+2)\sigma/\sigma_0}\right],
	\label{eqn:U1}
\end{equation}
where $\alpha$ is a constant given by 
%
\begin{equation}
\alpha=\frac{d(d-1)(d+2)}{2(d+4)\kappa^2\;{b_0}^2}. 
\end{equation}

This potential has the minimum zero at $\sigma=0$, provided that 
$d\ge 2$. Thus, in this case, the $4$-dimensional cosmological
constant is actually zero and $b_0$ is the present value of $b$. 
If  $d=1$ then the potential is completely flat. Hence, hereafter, we
assume that $d\ge 2$.

It is notable that the radion potential has been determined uniquely
up to the present value $b_0$ of the compactification radius
$b$. Since $b_0$ can be eliminated from all equations of $\chi^{lm}$
by redefinition of variables, all information about the corrected
radion potential has been obtained.

Finally, the perturbed Einstein-Hilbert action corrected by the
Casimir effect is given by Eq.~(\ref{eqn:perturbedEH2}), provided that 
$U_0$ is replaced by $U_1$ given by Eq.~(\ref{eqn:U1}). Therefore,
hereafter, we investigate the dynamics of the field $\chi^{lm}$
coupled to the radion field $\sigma$ in the expanding universe. These
fields are described by the following action.
%
\begin{equation}
I = I_{\sigma} + I_{\chi},
\end{equation}
where
%
\begin{eqnarray}
I_{\sigma} & = & -\int d^4x\sqrt{-\bar{g}^{ (0)}}\left[
\frac{1}{2}g^{\mu\nu}\partial_{\mu}\sigma\partial_{\nu}\sigma
+U_1(\sigma)\right],\nonumber\\
I_{\chi} & = & -\frac{1}{2}\int d^4x\sqrt{-\bar{g}^{ (0)}}\left[
e^{-4\frac{\sigma}{\sigma_0}}
g^{\mu\nu}\partial_{\mu}\chi^{lm}\partial_{\nu}\chi^{lm}
+\Lambda^{(T)}_{lm}\chi^{lm}\chi^{lm}\right].
\end{eqnarray}

Here, we mention that in the total action there may be other terms
which are second order in $\chi^{lm}$. In fact, there should appear
other mass terms for $\chi^{lm}$ since some zero modes may couple to
$\chi^{lm}$. However, providing the assumption (vii) in the next
section, these additional mass terms are small enough since zero modes
are considered as usual low-energy matter whose energy density and
pressure are bounded from above by $4$-dimensional cosmological
parameters. Therefore, the above action is enough for our purpose.


\section{Parametric resonance}
	\label{eqn:resonance}

In this section we investigate parametric resonance of the field
$\chi^{lm}$ due to the oscillation of the radion field $\sigma$ in the
expanding universe. For this purpose, we specify $4$-dimensional
geometry and initial condition somewhat. 
\begin{itemize}
 \item [(iv)] The background $4$-dimensional geometry is expressed by
 the flat FRW metric, and the radion field is homogeneous in the FRW
 universe. 
 \item [(v)] Initially, there is no excitation of Kaluza-Klein
 modes. Thus, the FRW universe is driven by the radion field as well
 as by usual homogeneous matter corresponding to zero modes.
\end{itemize}

Since the radion potential $U_1(\sigma)$ has exponential terms, it
seems difficult to analyze oscillation of $\sigma$ with large
amplitude analytically. Hence, we assume that the amplitude of the
oscillation is small enough. Moreover, for further simplicity, we
assume that the time scale of the cosmological expansion is much
longer than the time scale of the radion oscillation. Namely, we
assume the following. 
\begin{itemize}
 \item [(vi)] We assume that the amplitude of the radion oscillation
 around $\sigma=0$ is small compared with $\sigma_0$. 
 \item [(vii)] The energy scale of the cosmological expansion is much 
 lower than that of compactification: we assume that $Hb_0\ll 1$ and
 $\dot{H}b_0^2\ll 1$, where $H\equiv\dot{a}/a$ and the dot denotes the 
 derivative with respect to the cosmological time. 
\end{itemize}
Note that the assumption (vi) leads to the so called narrow resonance
regime for the parametric resonance.

\subsection{FRW background and homogeneous radion}

According to the assumptions (iv) and (v), we consider the homogeneous 
radion field and the flat FRW universe, which is driven by the radion
field and usual matter corresponding to zero modes:
%
\begin{eqnarray}
g_{\mu\nu}dx^{\mu}dx^{\nu} & = & -dt^2+{a(t)}^2(dx^2+dy^2+dz^2),
	\nonumber\\
\sigma & = & \sigma (t).
\end{eqnarray}
The cosmological evolution equations are
%
\begin{eqnarray}
 H^2 & = &\frac{8\pi G_N}{3}(\rho_{\sigma}+\rho_0)
	\nonumber\\
 \dot{H} + H^2 & = & -\frac{4\pi G_N}{3}
	\left[(\rho_{\sigma}+\rho_0)
	+3(p_{\sigma}+p_0)\right],
\end{eqnarray}
where the subscripts '$\sigma$' and '$0$' mean contributions from the 
radion $\sigma$ and those from usual matter corresponding to zero
modes, respectively. Here, $H\equiv\dot{a}/a$, $G_N=\kappa^2/8\pi$,
and the dot denotes derivative with respect to the time $t$.

Now, because of the symmetry of the background, it is convenient to
analyze the field $\chi^{lm}$ in the momentum space. Hence, we
decompose the field $\chi^{lm}$ as follows:
%
\begin{equation}
\chi_{lm} = a^{-\frac{3}{2}} \int
\frac{\sqrt{2}\;d^3\mbox{\boldmath$k$}}{(2\pi)^3}
[\chi^{(1)}_{lm\mbox{\boldmath$k$}}(t)
\cos(\mbox{\boldmath$k$}\cdot\mbox{\boldmath$x$})
+\chi^{(2)}_{lm\mbox{\boldmath$k$}}(t)
\sin(\mbox{\boldmath$k$}\cdot\mbox{\boldmath$x$})],
\end{equation}
where $\chi^{(1,2)}_{lm\mbox{\boldmath$k$}}$ are real functions of the 
cosmological time $t$ and the vector $\mbox{\boldmath$x$}$ represents
$(x,y,z)$. Correspondingly, the action $I_{\chi}$ is rewritten as
%
\begin{eqnarray}
& &I_{\chi}=\int dt L_{\chi},\nonumber\\
& &L_{\chi}=\frac{1}{2}\sum_{l, m}\sum_{i=1,2}
	\int d^3\mbox{\boldmath $k$}
         e^{-4\frac{\sigma}{\sigma_0}}\left[\left\{
	\partial_t\;\chi^{(i)}_{lm\mbox{\boldmath$k$}}(t)\right\}^2
	-m^2(t)\left\{\chi^{(i)}_{lm\mbox{\boldmath$k$}}(t)\right\}^2
	\right],\label{eqn;31}
\end{eqnarray}
where 
%
\begin{equation}
m^2(t)=\Lambda^{(T)}_{lm}+e^{-4\frac{\sigma}{\sigma_0}}
\frac{\mbox{\boldmath$k$}\cdot\mbox{\boldmath$k$}}{a^2}-
\left\{\frac{4}{9}H^2+\frac{3}{2}\dot{H}
-6H\dot{\left(\frac{\sigma}{\sigma_0}\right)}\right\}.
\end{equation}

Although the kinetic term in the above Lagrangian has time dependence, 
we can eliminate it by using a new time variable $\tau$ defined by
%
\begin{equation}
\frac{d\tau}{dt}=\exp{\left[4\frac{\sigma(t)}{\sigma_0}\right]}.
	\label{eqn:dtaudt}
\end{equation}
In fact, the equations of motion for
$\chi^{(1,2)}_{lm\mbox{\boldmath$k$}}$ are written as 
%
\begin{equation}
\frac{\partial^2Q}{\partial{{\tau}^2}}+{\Omega}^2(\tau)Q =0,
	\label{eqn:EOM-Q}
\end{equation}
where $Q$ denotes $\chi^{(1,2)}_{lm\mbox{\boldmath$k$}}$ and $\Omega$
is defined by 
%
\begin{equation}
\Omega^2(\tau)= m^2(t(\tau))
	\exp{\left[-4\frac{\sigma(t(\tau))}{\sigma_0}\right]}.
\end{equation}
The hamiltonian with respect to the time $\tau$ is 
%
\begin{equation}
{\cal H}\equiv\frac{f}{2}\left(P^2+{\Omega}^2Q^2\right), 
	\label{eqn:H-Q}
\end{equation}
where $P$ is the momentum conjugate to $Q$.

After the oscillation of $\sigma$ ends up at $\sigma=0$, the right
hand side of Eq.~(\ref{eqn:dtaudt}) becomes unity. Thus, after the end
of the oscillation, the hamiltonian ${\cal H}$ with respect to $\tau$
coincides with the hamiltonian with respect to the original
cosmological time $t$. At the same time, those hamiltonians become
conserved quantities. Hence, we can analyze the excitation of
$\chi^{(1,2)}_{lm\mbox{\boldmath$k$}}$ by using the new time variable
$\tau$.

\subsection{Small oscillation of the radion}

Providing the assumptions (vi) and (vii), the equation of motion for
$\sigma$ gives  
%
\begin{equation}
\frac{\sigma(t)}{\sigma_0}=\tilde{\sigma}(t)\cos\omega(t-t_0),
	\label{eqn:sigma/sigma0}
\end{equation} 
where $t_0$ is a constant, $\omega$ is defined by
%
\begin{equation}
\omega=\frac{1}{2}\sqrt{U_1''(0)} 
	=\frac{1}{b_0}\sqrt{\frac{d-1}{2}},
\end{equation}
and $\tilde{\sigma}(t)$ ($\ll 1$) is a slowly varying amplitude:
%
\begin{equation}
\tilde{\sigma}(t)\propto a^{-3/2}. 
\end{equation}

Now, by using Eq.~(\ref{eqn:sigma/sigma0}), we can express
$\Omega^2$ in terms of the new time variable $\tau$ up to the first
order in $\tilde{\sigma}$:  
%
\begin{equation}
{\Omega}^2(\tau)={\omega}^2\left[
	A+B+\tilde{\epsilon}\cos\{2\omega(\tau-\tau_0)\}
	+O(\tilde{\epsilon}^2)	\right],\label{eqn:Omega-tau}
\end{equation}
where $\tau_0$ is a constant and 
%
\begin{eqnarray}
 A & = &
	\frac{2{b_0}^2(\mbox{\boldmath$k$}^2/a)}{d-1}
	-\omega^2\left(\frac{9}{4}H^2+\frac{3}{2}\dot{H}\right),
	\label{eqn;50}\\
 B & = & 
	\frac{2(d+4)\left\{l(l+d-1)+(d-2)(d-1)\right\}-d(d-1)(d+2)}
	{b_0^2\;{\omega}^2},\\
 \tilde{\epsilon} & = &
	\tilde{\sigma}\left\{(d-1)\frac{d(d+2)(d+8)-l(l+d-1)(d-2)(d+10)}
	{b_0^2{\omega}^2}
	-\frac{8}{{\omega}^2}(\mbox{\boldmath$k$}^2/a)^2+4(d+4)\right\}.
\end{eqnarray}
The first term in $A$ is essentially the squared momentum along the
three-dimensional spatial directions divided by $\omega^2$, $B$ is
essentially the squared momentum along $S^d$ divided by $\omega^2$,
and $\tilde{\epsilon}$ is the properly normalized amplitude of the 
radion oscillation.

\subsection{Result}

The expression (\ref{eqn:Omega-tau}) is of almost the same form as the 
expression of $\Omega^2$ investigated in Ref.~\cite{Mukohyama}. Hence,
we can analyze the parametric resonance in the same way as in 
Ref.~\cite{Mukohyama}. Thus, the necessary and sufficient condition
for the catastrophic creation of quanta of the Kaluza-Klein mode is 
%
\begin{eqnarray}
 B & \le & 1,\nonumber\\
 |B-1| & = & O(\tilde{\epsilon}). 
\end{eqnarray}
However, it is easily shown that 
%
\begin{equation}
 B > 3
\end{equation}
for any values of $d$ ($\ge 2$) and $l$ ($\ge 2$). (See the setting
(c) in subsection~\ref{subsec:assumptions}.)
Therefore, it is concluded that quanta of the field $\chi^{lm}$ are
not overproduced by the parametric resonance due to the small
oscillation of the radion $\sigma$.

\section{Summary and Discussions}
	\label{sec:summary}

We have investigated excitation of Kaluza-Klein modes due to the
parametric resonance caused by oscillation of radius of
compactification in the narrow resonance regime. In particular, we
have considered a gravitational perturbation around a $D$-dimensional
spacetime compactified on a ($D-4$)-sphere. Among many Kaluza-Klein
modes of the gravitational perturbation we have investigated
part of Kaluza-Klein modes which correspond to massive scalar fields
in $4$-dimension. We have obtained that quanta of these modes are not
excited so catastrophically in the narrow resonance regime.

Now, several comments are in order.

First, we discuss about the consistency among the assumption (v), (vi) 
and (vii). Energy density $\rho_{\sigma}$ and pressure $p_{\sigma}$
of the radion is of the following order. 
%
\begin{equation}
 \rho_{\sigma} \sim p_{\sigma} \sim 
	\tilde{\epsilon}^2b_0^{-2}\kappa^{-2},
\end{equation}
where $\tilde{\epsilon}\sim\sigma/\sigma_0$. Thus, the condition
$H^2\sim\kappa^2\rho_{\sigma}$ can be satisfied, even if
$\tilde{\epsilon}\ll 1$, $Hb_0\ll 1$ and $\dot{H}b_0^2\ll 1$. In other 
words, the three assumptions (v), (vi) and (vii) can be satisfied
simultaneously.

Next, we discuss about validity of our strategy. Although we have
investigated only a particular class of Kaluza-Klein modes in this
paper, in principle we have to investigate all Kaluza-Klein modes in
order to judge whether a way of compactification is acceptable or
not. If we find that one of these many Kaluza-Klein modes is excited
abundantly, then the way of compactification should be rejected. In
this respect, there may be an embarrassing expectation: it is easily
expected that, even if we find that $N_{KK}\gg 1$, the result will be
significantly changed by backreaction, where $N_{KK}$ is the number of
created quanta of a Kaluza-Klein mode. In fact, by the energy
conservation and the assumptions (v) and (vii), it is easily concluded
that $N_{KK}\ll 1$. Thus, it might be expected that under the
assumptions (v) and (vii) we might not be able to judge whether a way
of compactification is acceptable or not. However, this embarrassing 
consideration does {\it not} spoil our analysis. Actually, if we
obtain $N_{KK}\gg 1$ by ignoring the backreaction, then this
overestimating result implies at least that $\rho_{KK}$ is not
negligible compared with $\rho_{\sigma}$ since the backreaction
becomes effective only when $\rho_{KK}$ is not negligible compared
with $\rho_{\sigma}$, where $\rho_{KK}$ is energy density of the
Kaluza-Klein mode. Therefore, providing the assumption (v), the result
$N_{KK}\gg 1$ obtained by ignoring the backreaction implies that
$\rho_{KK}$ is not negligible compared with energy density of
radiation, which should be rejected as explained in
Sec.~\ref{sec:introduction}.  
Because of almost the same reason, we can justify the assumption
(iii). First, provided that the higher-order coupling constant is at
most of order unity in the unit of $b_0$, those higher-order terms
should be inefficient as long as $\rho_{KK}\ll b_0^{-4}$. Physically,
this statement is almost equivalent to the statement that sufficiently 
small oscillation does not feel any higher-order corrections to its
potential. Therefore, we can neglect higher-order terms if the created 
number of Kaluza-Klein quanta is small. (See the assumption (v) for
the initial condition.) In other words, if higher-order terms 
becomes efficient at some epoch, then it implies that the number of
Kaluza-Klein quanta created so far is not small, and such a situation 
should be rejected as explained in Sec.~\ref{sec:introduction}.

Finally, we comment on possible future works. 
(A) We have to investigate Kaluza-Klein modes of other fields as well
as other gravitational modes. 
(B) We have to take inflation into consideration. 
(C) If we can extend our analysis to the broad resonance regime, then
it will give stronger constraints on ways of compactification. Note
that, even in the broad resonance regime, Eqs.~(\ref{eqn:EOM-Q}) and
(\ref{eqn:H-Q}) can be used. 
(D) Other mechanisms of compactification might deserve considering,
for example, the monopole-like configuration of an anti-symmetric
field~\cite{monopole}. 
(E) It might be also interesting to investigate the parametric
resonance of Kaluza-Klein modes in the Randall-Sundrum
scenario~\cite{RS} with stabilized radion~\cite{radion-brane}.

\begin{acknowledgments}
KU would like to thank Professor H. Kodama, Professor M. Sasaki, 
Professor M. Sakagami, Professor J. Soda, S. Tsujikawa, K. Koyama 
for discussions.
KU would also like to thank K. Yamashita for continuing encouragement.
YM would like to thank Professor T. Nakamura for continuous encouragement.
SM would like to thank Professor W. Israel for continuing
encouragement and Professor L. Kofman for his warm hospitality in
Canadian Institute for Theoretical Astrophysics. SM's work is
supported by the CITA National Fellowship and the NSERC operating
research grant. 

\end{acknowledgments}


\appendix


\section{Second-order Einstein action}\label{sec:action}

The purpose of this appendix is to expand the Einstein-Hilbert action
by $h_{MN}$ up to the second order.
Although in the main part of this paper all quantities defined in
$D$-dimension have ``$\bar{\ }$'', we omit it in this for simplicity. 

Let us decompose the metric $g_{MN}$ into the background
$g_{MN}^{(0)}$ and perturbation $h_{MN}$: 
%
\begin{equation}
g_{MN} =g_{MN}^{(0)}+h_{MN}. 
\end{equation}
Hereafter, we use the background metric $g_{MN}^{(0)}$ to raise the
indices of $h_{MN}$: 
%
\begin{eqnarray}
 {h^M}_N & = & {h_N}^M\equiv g^{MO}h_{ON},\nonumber\\
 h^{MN} & \equiv & g^{(0)MO} g^{(0)NP}h_{OP},\nonumber\\
 h & \equiv & g^{(0)MN}h_{MN}.
\end{eqnarray}

First, we can easily expand $g^{MN}$, $\sqrt{-g}$ and the Christoffel
symbol $\Gamma^P_{MN}$ as follows, where $g$ is the determinant of
$g_{MN}$. 
%
\begin{eqnarray}
 g^{MN} & = & g^{MN\;(0)}-h^{MN}+h^{MO}{h_O}^N+O(h^3),\nonumber\\
 \sqrt{-g} & = & 
	\sqrt{-g^{(0)}}\left\{1+\frac{1}{2}h+\frac{1}{2}
	\left(\frac{1}{4}h^2-\frac{1}{2}\;h^{MN}h_{MN}\right) 
	\right\}+O{(h^3)}, \nonumber\\
 \Gamma^P_{MN} & = & 
	{\Gamma^{(0)}}^P_{MN}+{\Gamma^{(1)}}^P_{MN}+
	{\Gamma^{(2)}}^P_{MN}+O(h^3),
\end{eqnarray}
where $g^{(0)}$ and ${\Gamma^{(0)}}^P_{MN}$ are the determinant and
the Christoffel symbol of the background metric $g_{MN}^{(0)}$, and 
%
\begin{eqnarray}
 {\Gamma^{(1)}}^P_{MN} & \equiv & 
	\frac{1}{2}\;{g^{(0)}}^{PQ}(h_{QM\:,\:N}
	+h_{QN\:,\:M}-h_{MN\:,\:Q}),\nonumber\\
 {\Gamma^{(2)}}^P_{MN} & \equiv & 
	-\frac{1}{2}\;h^{PQ}(h_{QM\:,\:N}
	+h_{QN\:,\:M}-h_{MN\:,\:Q}).
\end{eqnarray}
Here, `` , '' denotes the partial derivative.

Next, the Ricci tensor is expanded as follows.
%
\begin{eqnarray}
 R_{MN} & = & R^P_{MPN}=\Gamma^P_{MN\:,\:P}
	-\Gamma^P_{MP\:,\:N}+\Gamma^P_{SP}
	\Gamma^S_{MN}-\Gamma^P_{SN}
	\Gamma^S_{MP}\nonumber\\ 
 & = & R^{(0)}_{MN}+R^{(1)}_{MN}+R^{(2)}_{MN}+O(h^3), 
\end{eqnarray}
where $R^{(0)}_{MN}$ is the Ricci tensor for the background metric
$g_{MN}^{(0)}$, and
%
\begin{eqnarray}
 R^{(1)}_{MN} & = & {\Gamma^{(1)}}^P_{MN\:,\:P}
	-{\Gamma^{(1)}}^P_{MP\:,\:N} ,\\
 R^{(2)}_{MN} & = & {\Gamma^{(2)}}^P_{MN\:,\:P}
	-{\Gamma^{(2)}}^P_{MP\:,\:N}
        +{\Gamma^{(1)}}^P_{SP}{\Gamma^{(1)}}^S_{MN}
	-{\Gamma^{(1)}}^P_{SN}{\Gamma^{(1)}}^S_{MP}.
\end{eqnarray}

Therefore, we obtain the following expansion.
%
\begin{equation}
 \sqrt{-g}R = \sqrt{-g}g^{MN}R_{MN}
	= \sqrt{-g^{0}}
	\left(R^{(0)}+R^{(1)}+R^{(2)}+O(h^3)\right),
\end{equation}
where $R^{(0)}$ is the Ricci scalar for the background metric
$g_{MN}^{(0)}$, and
%
\begin{eqnarray}
 R^{(1)} & \equiv & \frac{1}{2}h\;R^{(0)}-h^{MN}R^{(0)}_{MN}
	+{g^{(0)}}^{MN}{R^{(1)}}_{MN}\nonumber\\
	& = & 
	-h^{MN}\left(R^{(0)}_{MN}-\frac{1}{2}R^{(0)}g_{MN}\right)
	+({h^{MN}}_{;N}-h^{;M})_{;M},
	\nonumber\\ 
 R^{(2)} & \equiv & \frac{1}{2}\left(\frac{1}{4}h^2-\frac{1}{2}h^{PQ}
	h_{PQ}\right)R^{(0)}+h^{MP}{h_P}^NR^{(0)}_{MN}
	+{g^{(0)}}^{MN}{R^{(2)}}_{MN}\nonumber\\
	& & -\frac{1}{2}\;h\;h^{MN}R^{(0)}_{MN}
	-h^{MN}R^{(1)}_{MN}+\frac{1}{2}h\;{g^{(0)}}^{MN}R^{(1)}_{MN}
	\nonumber\\
	& = & \frac{1}{8}(h^2-2h^{PQ}h_{PQ})R^{(0)}+
	\frac{1}{2}(2h^{MP}{h_P}^N-hh^{MN})R^{(0)}_{MN}
	\nonumber\\
	& & +\frac{1}{4}\left\{{h^{MN}}_{;P}+\left(2{h^P}_{M\;;\;N}
	 -{h_{MN}}^{;\;P}\right)+h_{;\;M}\left(h^{;\;M}-
	2{h^{MN}}_{;\;N}\right)\right\} + X^P_{\ ;P}.
\end{eqnarray}
Here, $X^P$ is a vector constructed from $h_{MN}$, and `` ; '' denotes
the covariant derivative compatible with the background metric
$g_{MN}^{(0)}$.

Finally, by using the above expressions and integrating by part, we
obtain the perturbed Einstein-Hilbert action up to the second order in
$h_{MN}$. 
%
\begin{eqnarray} 
 I_{EH} & \equiv & \frac{1}{2\kappa^2} 
 	\int d^Dx\sqrt{-g}( R^{(0)} -2\Lambda)\nonumber\\
 & =  & \frac{1}{2\kappa^2} 
 	\int d^Dx\sqrt{-g^{(0)}} \left[ R^{(0)} - 2\Lambda
	-h^{MN}\left(R^{(0)}_{MN}-\frac{1}{2}R^{(0)}g^{(0)}_{MN}
	+\Lambda g^{(0)}_{MN}\right)\right.\nonumber\\
 & & + \frac{1}{8}\left( h^2-2h^{MN}h_{MN}\right)R^{(0)} 
	+\frac{1}{2}\left(2h^{MM'}h_{M'}^N-hh^{MN}\right)
	R^{(0)}_{MN}  \nonumber  \\ 
 & &  +\frac{1}{4} \left\{ h^{MN}_{\quad;M'}
	\left(2h^{M'}_{M;N}- h_{MN}^{\quad;M'}\right) 
	+ h_{;M} \left(h^{;M}-2h^{MN}_{\quad;N}\right) 
	\right\}\nonumber\\
 & & \left.
	-\Lambda\left(\frac{1}{4}h^2-\frac{1}{2}h^{MN}h_{MN} 
	\right)  + O(h^3)\right].
\end{eqnarray}


\section{Harmonics on $d$-sphere}
	\label{sec:harmonics}

In this Appendix we give definitions and basic properties of scalar,
vector and tensor harmonics on a unit
$d$-sphere~\cite{harmonics}. Throughout this Appendix we will use the
notation that $\Omega_{ij}$ is the metric of the unit $d$-sphere and
$D$ is the covariant derivative compatible with $\Omega_{ij}$.

\subsection{scalar harmonics}

The scalar harmonics is supposed to satisfy the following relations.
%
\begin{eqnarray}
 D^2 Y_{lm} + l(l+d-1)Y_{lm} & = & 0\qquad (l\geq0),\\
 \int d^dx{\sqrt{\Omega}}\;
	Y_{lm} Y_{l'm'} & = & {\delta}_{ll'}\;{\delta}_{mm'}.
\end{eqnarray}

\subsection{vector harmonics}

First, in general, a vector field $V$ on the unit $d$-sphere can be
decomposed as
%
\begin{equation} 
 V_i=V_{(T)\:i}+\partial_i f , 
\end{equation}
where $f$ is a function and $V_{(T)}$ is a transverse vector field:
%
\begin{equation}
D^i (V_{(T)})_i=0 .
\end{equation}

Thus, the vector field $V_i$ can be expanded by using the scalar
harmonics $Y_{lm}$ and transverse vector harmonics $V_{(T)\;lm}$ as 
%
\begin{equation}
 V_i = \sum_{lm}\left[
	c_{(T)}^{lm}(V_{(T)\;lm})_i+c_{(L)}^{lm}\partial_i Y_{lm}
	\right],\label{eqn:dY+V}
\end{equation}
where $c_{(T)}^{lm}$ and $c_{(L)}^{lm}$ are constants, and the
transverse vector harmonics $V_{(T)\;lm}$ ($l\ge 1$) is supposed to
satisfy the following relations.
%
\begin{eqnarray}
 D^2\;V_{(T)\;lm} + \left\{l(l+d-1)-1\right\}V_{(T)\;lm} & = & 0,
	\nonumber\\
 D^i(V_{(T)\;lm})_i & = & 0,\nonumber\\
\int d^dx{\sqrt{\Omega}}\;\Omega^{ij}(V_{(T)\;lm})_i\;(V_{(T)\;l'm'})_j
	& = & {\delta}_{ll'}\;{\delta}_{mm'}.
\end{eqnarray}
>From Eq.~(\ref{eqn:dY+V}), it is convenient to define longitudinal
vector harmonics $V_{(L)\:lm}$ by 
%
\begin{equation}
 (V_{(L)\:lm})_i=\partial_i Y_{lm}\qquad (l\geq 1).
\end{equation}

It is easily shown that the longitudinal vector harmonics satisfies the 
following properties. 
%
\begin{eqnarray}
 D^2V_{(L)\: lm} + \{l(l+d-1)-(d-1)\}V_{(L)\:lm} & = & 0,\nonumber\\ 
 D^i(V_{(L)lm})_i & = & -l(l+d-1)Y_{lm},\nonumber\\
 D_{[i}(V_{(L)\: lm})_{j]} & = & 0,\nonumber \\
 \int d^dx{\sqrt{\Omega}}\;{\Omega}^{ij}
	(V_{(L)\:lm})_i (V_{(L)\:l'm'})_j
	& = & l(l+d-1){\delta}_{ll'}\;{\delta}_{mm'},\nonumber\\
 \int d^dx {\sqrt{\Omega}}\;{\Omega}^{ij}(V_{(T)\:lm})_i 
	(V_{(L)\:l'm'})_j & = & 0.  
\end{eqnarray}

\subsection{Tensor harmonics}

First, in general, a symmetric second-rank tensor field $T_{ij}$ can
be decomposed as
%
\begin{equation}
 T_{ij}=T_{(T)\;ij} + D_i\;V_j+D_j\;V_i + f\Omega_{ij},
\end{equation}
where $f$ is a function, $V_i$ is a vector field and $T_{(T)\;ij}$ is
a transverse traceless symmetric tensor field:
%
\begin{eqnarray}
 (T_{(T)})^i_i & = & 0,\nonumber\\
 D^i (T_{(T)})_{ij} & = &0. 
\end{eqnarray}

Thus, the tensor field $T_{ij}$ can be expanded by using the vector
harmonics $V_{(T)\;lm}$ and $V_{(L)\:lm}$, and transverse vector
harmonics $V_{(T)\;lm}$ as 
%
\begin{eqnarray}
 T_{ij} & = & \sum_{lm}\left[
	c_{(T)}^{lm}(T_{(T)\;lm})_{ij}
	+ c_{(LT)}^{lm}\left\{ 
	D_i\;(V_{(T)\;lm})_j+D_j\;(V_{(T)\;lm})_i \right\}
	\right.\nonumber\\
 & & \left.+ c_{(LL)}^{lm}\left\{ 
	D_i\;(V_{(L)\;lm})_j+D_j\;(V_{(L)\;lm})_i \right\}
	+ c_{(Y)}^{lm} Y_{lm}\Omega_{ij}\right],
	\label{eqn:dV+T}
\end{eqnarray}
where $c_{(T)}^{lm}$, $c_{(LT)}^{lm}$, $c_{(LL)}^{lm}$ and
$c_{(Y)}^{lm}$ are constants, and the transverse tensor harmonics
$T_{(T)\;lm}$ ($l\ge 2$) is supposed to satisfy the following
relations. 
%
\begin{eqnarray}
 D^2 T_{(T)\: lm}+\{l(l+d-1)-2\}T_{(T)\: lm} & = & 0,\nonumber\\ 
 (T_{(T)\: lm})^i_i & = & 0,\nonumber\\
 D^i (T_{(T)\: lm})_{ij} & = &0,\nonumber\\
 \int d^dx{\sqrt{\Omega}}\;{\Omega}^{ii'}\;{\Omega}^{jj'}
	(T_{(T)\: lm})_{ij} (T_{(T)\: l'm'})_{i'j'} 
	& = & {\delta}_{ll'}{\delta}_{mm'}.
\end{eqnarray}
From Eq.~(\ref{eqn:dV+T}), it is convenient to define tensor harmonics
$T_{(LT)\:lm}$, $T_{(LL)\:lm}$, and $T_{(Y)\:lm}$ by 
%
\begin{eqnarray}
 (T_{(LT)\: lm})_{ij} & \equiv &  D_i (V_{(T)\:lm})_j +  
	D_j (V_{(T)\: lm})_i,\qquad (l\ge 2)\nonumber\\
 (T_{(LL)\: lm})_{ij} & \equiv & D_i (V_{(L)\:lm})_j 
	+D_j (V_{(L)\: lm})_i - \frac{2}{d} \Omega_{ij}
	D^k(V_{(L)\:lm})_k,\qquad (l\ge 2) \nonumber\\
 (T_{(Y)\;lm})_{ij} & \equiv & \Omega_{ij}Y_{(Y)\;lm}
	\qquad (l\ge 0).
\end{eqnarray}

It is easily shown that these tensor harmonics satisfy the following
properties.
%
\begin{eqnarray}
 D^2 T_{(LT)\: lm} + \{l(l+d-1)-(d+2)\}T_{(LT)\:lm} & = & 0
	,\nonumber\\ 
 D^i (T_{(LT)\: lm})_{ij} & = & -\{l(l+d-1)-d\}(V_{(T)\:lm})_j
	 ,\nonumber\\
 (T_{(LT)\: lm})^i_i & = & 0,
\end{eqnarray}
%
\begin{eqnarray}
 D^2 T_{(LL)\: lm} + \{l(l+d-1)-2d\}T_{(LL)\:lm} & = &0 
	,\nonumber\\
 D^i(T_{(LL)\: lm})_{ij} & = & -\frac{2(d-1)}{d}\{l(l+d-1)-d\}
	(V_{(L)\:lm})_{j},\nonumber\\
 {(T_{(LL)\: lm})^i}_i & = & 0,
\end{eqnarray}
and 
%
\begin{eqnarray}
 D^2 T_{(Y)\;lm} + l(l+d-1)\;T_{(Y)\;lm} & = & 0,\nonumber\\
 D^i (T_{(Y)\;lm})_{ij} & = & (V_{(L)\:lm})_j,\nonumber\\ 
 (T_{(Y)\;lm})^i_i & = & d\;Y_{lm}.
\end{eqnarray}

It is also easy to show the following formulas of integral as
well as the orthogonality between different types of tensor harmonics. 
%
\begin{eqnarray}
 \int d^dx{\sqrt{\Omega}}\;{\Omega}^{ii'}\;
	{\Omega}^{jj'}(T_{(LT)\: lm})_{ij}(T_{(LT)\:l'm'})_{i'j'}
 & = & 2\{l(l+d-1)-d\}{\delta}_{ll'}\;{\delta}_{mm'},\nonumber\\
 \int d^dx{\sqrt{\Omega}}\;{\Omega}^{ii'}\;{\Omega}^{jj'} 
	(T_{(LL)\: lm})_{ij}(T_{(LL)\:l'm'})_{i'j'}
 & = & \frac{4l(l+d-1)(d-1)}{d}\{l(l+d-1)-d\}
	{\delta}_{ll'}\;{\delta}_{mm'},\nonumber\\ 
 \int d^dx{\sqrt{\Omega}}\;{\Omega}^{ii'}\;
	{\Omega}^{ii'} (T_{(Y)\;lm})_{ij}(T_{(Y)\;l'm'})_{i'j'}
 & = & d\;{\delta}_{ll'}\;{\delta}_{mm'}.
\end{eqnarray}

\subsection{Integration formulas}

Finally, we list up some other formulas of integration. 
%
\begin{eqnarray}
 \int d^dx{\sqrt{\Omega}}\;D^i(V_{(T)\: lm})_j\;D_i(V_{(T)\: l'm'})^j 
 & = & \{l(l+d-1)-1\}\delta_{ll'}\;\delta_{mm'},\nonumber\\
 \int d^dx{\sqrt{\Omega}}\;D^i(V_{(T)\: lm})_j\;D_i(V_{(T)\: l'm'})^j
 & = & -(d-1)\delta_{ll'}\;\delta_{mm'},\nonumber\\
 \int d^dx{\sqrt{\Omega}}\;D^i(T_{(T)\: lm})^{jk}
	D_i(T_{(T)\:l'm'})_{jk} 
 & = & \{l(l+d-1)-2\}\delta_{ll'}\;\delta_{mm'},\nonumber\\
 \int d^dx{\sqrt{\Omega}}\;D^i(T_{(Y)\: lm})^{jk}
	D_i(T_{(Y)\: l'm'})_{jk}
 & = & d\{l(l+d-1)-2\}\delta_{ll'}\;\delta_{mm'},\nonumber\\
 \int d^dx{\sqrt{\Omega}}\;D^k(T_{(T)\: lm})^{ij}
	D_i(T_{(T)\: l'm'})_{jk}
 & = & -d\delta_{ll'}\;\delta_{mm'},\nonumber\\
 \int d^dx{\sqrt{\Omega}}\;D^k(T_{(Y)\: lm})^{ij}
	D_i(T_{(Y)\: l'm'})_{jk}
 & = & l(l+d-1)\delta_{ll'}\;\delta_{mm'}.
\end{eqnarray}


\section{Gauge fixing}\label{sec:fix}

In this appendix we show that there is a particular choice of gauge in
which the gravitational perturbation $h_{MN}$ is expanded as
Eq.~(\ref{eqn:gaugefixedh}).

First, the infinitesimal gauge transformation of $h_{MN}$ due to a
vector field $\xi^M$ is expressed as
%
\begin{equation}
 h_{MN}\rightarrow 
	h_{MN}-\xi_{M;N}-\xi_{N;M},
	\label{eqn:gaugetr-h}
\end{equation}
where $\xi_{M}\equiv g^{(0)}_{MN}\xi^{N}$, and ``$;$'' denotes the
covariant derivative compatible with the background metric
$\bar{g}^{(0)}_{MN}$.

Next, we expand $h_{MN}$ and $\xi_{M}$ by harmonics on $S^d$, which are 
defined in Appendix~\ref{sec:harmonics}, as
%
\begin{eqnarray}
 h_{MN}dx^Mdx^N &=& \sum_{lm}
 	\left[ h_{\mu\nu}^{lm}
	Y_{lm}dx^{\mu}dx^{\nu}  
 	+ 2\{h_{(T)\mu}^{lm}(V_{(T)lm})_i
	+ h_{(L)\mu}^{lm}(V_{(L)lm})_i \}
	dx^{\mu}dx^i \right.\nonumber \\
 & & + \left.
	\{ h_{(T)}^{lm}(T_{(T)lm})_{ij}
	+ h_{(LT)}^{lm}(T_{(LT)lm})_{ij}
	+ h_{(LL)}^{lm}(T_{(LL)lm})_{ij}
	+ h_{(Y)}^{lm}(T_{(Y)lm})_{ij}\}
	dx^idx^j \right],\nonumber\\
\end{eqnarray}
and 
%
\begin{equation}
 \xi_M dx^M = \sum_{lm}\left[ \xi_{\mu}^{lm} Y_{lm}dx^{\mu}
 	+\{ \xi_{(T)}^{lm}(V_{(T)lm})_i+
	\xi_{(L)}^{lm}(V_{(L)lm})_i\}dx^i\right],
\end{equation}
Here, the coefficients $h_{\mu\nu}^{lm}$, $h_{(T)\mu}^{lm}$,
$h_{(L)\mu}^{lm}$, $h_{(T)}^{lm}$, $h_{(LT)}^{lm}$, $h_{(LL)}^{lm}$,
$h_{(Y)}^{lm}$, $\xi_{\mu}^{lm}$, $\xi_{(T)}^{lm}$ and
$\xi_{(L)}^{lm}$ depend only on the four-dimensional coordinates
$x^{\mu}$, while the harmonics depend only on the coordinates $x^i$ on
$S^d$.

Hence, by substituting the above expansions into
Eq.~(\ref{eqn:gaugetr-h}), we obtain the infinitesimal gauge
transformation of each coefficient. 
%
\begin{eqnarray}
 h_{\mu\nu}^{lm} & \rightarrow & 
	h_{\mu\nu}^{lm} -\left(\frac{b}{b_0}\right)^{-d}\left\{
	\nabla_{\mu}
	\left[\left(\frac{b}{b_0}\right)^{d/2}\xi_{\nu}^{lm}\right]
	+\nabla_{\nu}
	\left[\left(\frac{b}{b_0}\right)^{d/2}\xi_{\mu}^{lm}\right]
	\right\}
	+ d \left(\frac{b}{b_0}\right)^{-d/2}
	g_{\mu\nu}g^{\alpha\beta}\xi^{lm}_{\alpha}\partial_{\beta}
	\ln{\left(\frac{b}{b_0}\right)},
	\nonumber \\
 h_{(T)\mu}^{lm} & \rightarrow & 
	h_{(T)\mu}^{lm} -b^2
	\partial_{\mu}\left[b^{-1}\xi_{(T)}^{lm}\right],\nonumber \\
 h_{(L)\mu}^{lm} & \rightarrow &
	h_{(L)\mu}^{lm}-
	\left(\frac{b}{b_0}\right)^{-d/2}\xi_{\mu}^{lm}
	-b^2\partial_{\mu}\left[b^{-1}\xi_{(L)}^{lm}\right], 
	\nonumber\\
 h_{(T)}^{lm} & \rightarrow & h_{(T)}^{lm},
	\nonumber\\
 h_{(LT)}^{lm} & \rightarrow &  
	h_{(LT)}^{lm} - b\xi_{(T)}^{lm},\nonumber \\
 h_{(LL)}^{lm} & \rightarrow &  
	h_{(LL)}^{lm} - b\xi_{(L)}^{lm},\nonumber \\
 h_{(Y)}^{lm} & \rightarrow & 
	h_{(Y)}^{lm} - 2b^2\left(\frac{b}{b_0}\right)^{d/2}
	g^{\mu\nu}\xi^{lm}_{\mu}
	\partial_{\nu}\ln{\left(\frac{b}{b_0}\right)}
	+\frac{2l(l+d-1)}{d}b\xi_{(L)}^{lm},
\end{eqnarray}
where $\nabla$ is the covariant derivative compatible with the
$4$-dimensional metric $g_{\mu\nu}$.

Therefore, by choosing $\xi_{\mu}^{lm}$, $\xi_{(T)}^{lm}$ and
$\xi_{(L)}^{lm}$ as
%
\begin{eqnarray}
 \xi_{(T)}^{lm} & = & b^{-1}h_{(LT)}^{lm}\qquad (l\ge 2),\nonumber\\
 \xi_{(L)}^{lm} & = & b^{-1}h_{(LL)}^{lm}\qquad (l\ge 2),\nonumber\\
 \xi^{lm}_{\mu} & = & \left(\frac{b}{b_0}\right)^{d/2}h_{(L)\mu}^{lm}
	-b^2\left(\frac{b}{b_0}\right)^{d/2}
	\partial_{\mu}\left[b^{-1}\xi_{(L)}^{lm}\right]
	\qquad (l\ge 1),\nonumber\\
 \xi_{(L)}^{1m} & = & -\frac{1}{2b}h_{(Y)}^{1m}
	+b\left(\frac{b}{b_0}\right)^{d/2}
	g^{\mu\nu}\xi^{1m}_{\mu}\partial_{\nu}
	\ln{\left(\frac{b}{b_0}\right)}, 
\end{eqnarray}
we can always make a gauge transformation such that 
%
\begin{eqnarray} 
 h_{(LT)}^{lm} & \rightarrow & 0 \qquad (l\geq2),\nonumber\\
 h_{(LL)}^{lm} & \rightarrow & 0 \qquad (l\geq2),\nonumber\\
 h_{(L)\mu}^{lm} &\rightarrow&0 \qquad (l\geq1),\nonumber\\
 h_{(Y)}^{1m} & \rightarrow & 0.
\end{eqnarray}

Moreover, $h_{\mu\nu}^{00}$ and $h_{(Y)}^{00}$ can be eliminated by
redefining the $4$-dimensional metric $g_{\mu\nu}$ (or
$\hat{g}_{\mu\nu}$) and $4$-dimensional scalar field $b$ (or
$\sigma$). It can be easily shown that the gauge transformation of
$h_{\mu\nu}^{00}$ and $h_{(Y)}^{00}$ is equivalent to the
$4$-dimensional gauge transformation.

Therefore, we obtain the following expansion of $h_{MN}$:
%
\begin{equation}
 h_{MN}dx^Mdx^N = \sum_{lm}
	\left[ h_{\mu\nu}^{lm}Y_{lm}dx^{\mu}dx^{\nu}
	+ 2h_{(T)\mu}^{lm}(V_{(T)lm})_idx^{\mu}dx^i
	+ \{ h_{(T)}^{lm}(T_{(T)lm})_{ij}
	+ h_{(Y)}^{lm}(T_{(Y)lm})_{ij}\}dx^idx^j\right],
\end{equation}
where the summations are taken over $l\ge 1$ for the scalar and vector 
harmonics, and over $l\ge 2$ for the tensor harmonics.

The remaining gauge freedom is given by $\xi_{(T)}^{1m}$ and
$\xi_{\mu}^{00}$. The first one generates the gauge transformation of
$h_{(T)\mu}^{lm}$ as 
%
\begin{equation}
 (h_{(T)}^{1m})_{\mu} \rightarrow (h_{(T)}^{1m})_{\mu}
	-b^2\partial_{\mu}\left[ b^{-1}\xi^{1m}_{(T)}\right].
\end{equation}
Another remaining gauge freedom $\xi_{\mu}^{00}$ generates the usual
$4$-dimensional gauge transformation.


\end{document}